\documentclass[conference]{IEEEtran}
\IEEEoverridecommandlockouts
\usepackage[letterpaper, left=0.625in, right=0.625in, top=0.72in, bottom=1.02in]{geometry}
\usepackage{cite}
\usepackage{amsmath,amssymb,amsfonts,array}
\usepackage{algorithm,hyperref}
\usepackage{graphicx}
\usepackage{textcomp}
\usepackage{xcolor}
\usepackage{multirow}
\usepackage{xcolor}
\definecolor{MXY}{RGB}{252,8,235}
\usepackage{algpseudocode} % For a more structured pseudocode environment

\def\BibTeX{{\rm B\kern-.05em{\sc i\kern-.025em b}\kern-.08em
    T\kern-.1667em\lower.7ex\hbox{E}\kern-.125emX}}

    \makeatletter
\renewcommand\fs@ruled{\def\@fs@cfont{\bfseries}\let\@fs@capt\floatc@ruled
  \def\@fs@pre{\vspace{0.06in}\hrule height.8pt depth0pt \kern2pt}%
  \def\@fs@post{\kern2pt\hrule\relax}%
  \def\@fs@mid{\kern2pt\hrule\kern2pt}%
  \let\@fs@iftopcapt\iftrue}
\makeatother

\begin{document}

\title{Joint Association and Beamforming Design for Cooperative ISAC Systems}%\thanks{Identify applicable funding agency here. If none, delete this.}

% \author{\IEEEauthorblockN{1\textsuperscript{st} Abhishek Rajasekaran}
% \IEEEauthorblockA{\textit{Electrical and Computer Engineering} \\
% \and
% \and
% \and
% \and
% \and
\author{\IEEEauthorblockN{Abhishek Rajasekaran, Mehdi Karbalayghareh, Xiaoyan Ma, David J. Love, and Christopher G. Brinton}
\IEEEauthorblockA{{Electrical and Computer Engineering, Purdue University, IN 47907, USA} \\
Email:\{\href{mailto:rajasek1@purdue.edu}{rajasek1}, \href{mailto:mkarbala@purdue.edu}{mkarbala}, \href{mailto:ma946@purdue.edu}{ma946}, \href{mailto:djlove@purdue.edu}{djlove},
\href{mailto:cgb@purdue.edu}{cgb}\}@purdue.edu
}}

\maketitle

\begin{abstract}%Most existing works assume various simplifications like a single communication user (CU) or a single target or just one transmitting or receiving BS or do not assume a common waveform for both or do not consider the non-coherent radar case or assume designated BSs as the sensing or communication nodes, and thus do not establish the rigorous math for the actual system of interest that exploits the full potential of cooperative ISAC in terms of sensing and communication gain. 
% Cellular networks with integrated sensing and communication (ISAC) enable multiple base stations (BSs) to cooperate in both sensing and communication.(for communication and sensing) 
Cooperation among multiple base stations (BSs) improves both communication and sensing in integrated sensing and communication (ISAC) networks. However, it requires deciding which BSs serve each communication user (CU) and each target, and how each BS beamforms. In this work, we consider joint BS–CU and BS–target association, and beamforming for a general cooperative ISAC system, overcoming prior simplifications such as single-user/target setups, fixed BS roles, coherent radar or separate waveforms. We maximize a joint communication rate and position Cram\'{e}r-Rao lower bound (CRLB) based tradeoff utility while guaranteeing a minimum service per-CU using a data queue model. We develop a successive convex approximation (SCA)-based beamforming with closed-form updates. The discrete association problem is solved as an overarching layer using a low-complexity algorithm. Numerical results validate the effectiveness of the algorithm and unveil benefits of cooperative communication and sensing.
\end{abstract}

% \begin{IEEEkeywords}
% component, formatting, style, styling, insert
% \end{IEEEkeywords}

\section{Introduction}
Integrated sensing and communication (ISAC) is  recognized as one of the major enablers of sixth-generation (6G) wireless in the ITU-R (International Telecommunication Union-Radiocommunication Sector) recommendation~\cite{itu}. Specifically, ISAC helps achieve data transmission and target localization using shared infrastructure, resources, and waveforms, thus improving energy, spectral, and financial efficiency of the two capabilities~\cite{isacsurvey}. Motivated by its massive potential and numerous applications, industry and academia are working towards adopting ISAC as the global radio standard~\cite{6gbrinton,Nuria,nextgindustry}. 

To best exploit ISAC capabilities, cooperative/distributed architectures have gained interest, enabling multiple base stations (BSs) to collaborate as ISAC transceivers across cells and jointly enhance communication and sensing~\cite{coop1}. Such cooperation extends the coverage area and provides diverse sensing angles, leading to more accurate target localization~\cite{coopnotisac}. Recent work has shown that $N$ transceivers yield an enhanced average sensing performance with the $\ln^{2}{N}$ law~\cite{coop2}. On the communication side, coordinated multi-point (CoMP) transmission can enhance performance through joint transmission across BSs~\cite{coop1}. However, realizing these benefits requires addressing two major design problems: \emph{joint communication and sensing beamforming and association}, both of which must be effectively solved in real time. 
%\textcolor{blue}{[Mehdi: I would recommend adding a paragraph after this, describing the problem and challenges. to save space, we can cut the text in the related works (just a brief outline of the related literature)]}
%The first major problem of interest for this setup is joint communication and sensing beamforming. 

\subsection{Related Works}
Many works solve the beamforming problem for a single BS using communication-centric or sensing-centric approaches. For instance,~\cite{beam3} minimizes the orientation Cram\`er-Rao lower bound (CRLB) with a communication guarantee.~\cite{beam1} derives the beamformers for the cooperative multi-BS case for a single target using a sensing signal-to-interference-plus-noise ratio (SINR) objective with communication constraints, while~\cite{beam2} does it for a one target per-cell setup using a minimum communication SINR and CRLB tradeoff utility. All these works use semi-definite relaxation (SDR)-based techniques. 

On the other hand, incorporating the BS-communication user (CU) and BS-target associations introduces a mixed integer issue to the beamforming problem under consideration. For simpler system setups, existing approaches include exhaustive search with exponential complexity~\cite{assoc}, penalty-based binary relaxation~\cite{assoc,close2}, and large language model (LLM)-based association~\cite{close1}. Notably, binary relaxation introduces considerable complexity to the beamforming problem. %assumes each CU is served by one BS and~\cite{assoc} proposes an overarching exponential search or a penalty-based binary relaxation to overcome the integer issue for a simpler system setup.~\cite{close2} uses an large language model (LLM)-based method for association, while~\cite{close1} also solves using penalty-based binary relaxation, for simpler system setups.
Recently,~\cite{queue} introduced data traffic queue model to the setup, accounting for its stability in an edge-computing setup. 

Therefore, there is a clear research gap for a low-complexity joint beamforming and association algorithm that works for the most general setup of the cooperative ISAC problem with multiple BSs, CUs, and targets, while guaranteeing minimum service under queue limits.

\subsection{Outline of Methodology and Contributions}
To address this gap, we propose the maximization of the joint sum-rate and inverse CRLB based utility with data service, power, and association guarantees for a general multi-BS, cooperative joint CoMP communication and cooperative non-coherent detection setup. Further, we apply successive convex approximations (SCA) to the objective and derive closed-form updates for the communication and sensing beamformers, and the auxiliary variables introduced. An iterative algorithm is listed for the beamforming problem. We then propose to fix CUs associations based on distance, as concluded optimal in~\cite{coop2}. Sensing associations are dynamically updated as an overarching low-complexity layer after each beamforming update based on a CRLB based sensing utility.

Thus, we consider the most general setup of the cooperative ISAC system with a neutral communication-sensing tradeoff utility, propose a low-complexity alternative to penalty-based binary relaxation while extending the association problem to each BS-CU and BS-target, and derive a low-complexity SCA-based beamforming with closed-form updates, while guaranteeing a minimum service (communication rate) for each CU. 

\section{System Model}

\begin{figure}[t!]
    \centering
    \includegraphics[width=0.35\textwidth]{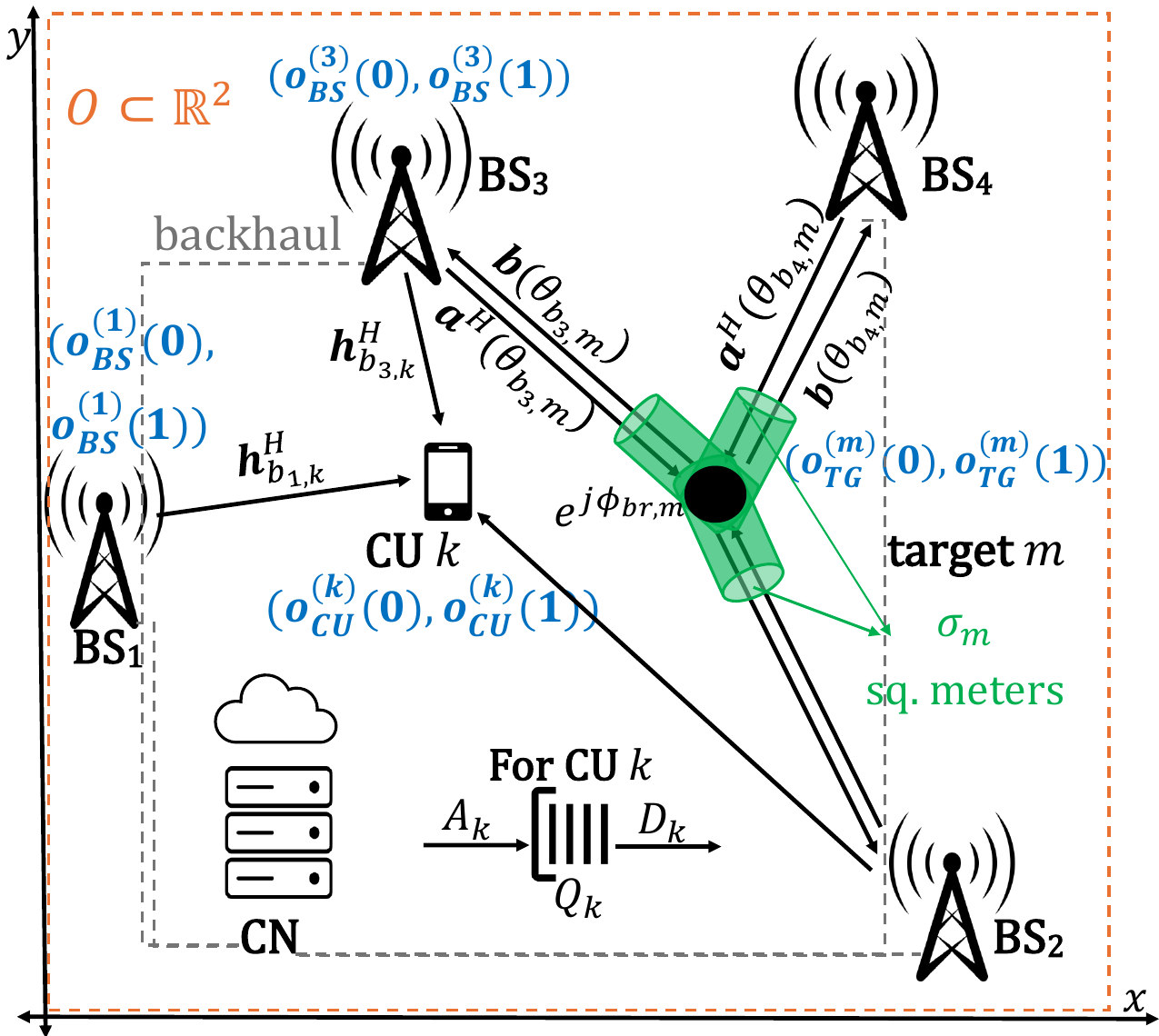}
    \caption[Cooperative ISAC system.]{% 
    Cooperative ISAC system.\protect\footnotemark}
    \label{fig:sysmodel}
            \vspace{-12pt}
\end{figure}
\footnotetext{The slot index \(t\) is dropped from the channels
and positions for readability.}

Consider an ISAC network in a bounded space $\mathcal{O} \subset \mathbb{R}^2$ with a fixed set of BSs $\mathcal{B}$ at known locations $\{\boldsymbol{o}^{(b)}_{\mathrm{BS}}\mid b \in \mathcal{B}\}\subset\mathcal{O}$. Each BS has $N_t$ transmit and $N_r$ receive antennas and is connected to the core network (CN) through wired backhaul. CN has access to communication channel state information (CSI), CU data queues, and predicted target positions. %Targets are assumed to have been initially detected and provided with coarse position estimates. %Their precise positions are to be estimated. 

Time is divided into slots of duration $\Delta$. Let $\mathcal{K}(t)$ and $\mathcal{M}(t)$ denote the active CUs and sensing targets in the current slot~$t$, respectively. Further, let $\{\boldsymbol{o}^{(k)}_{\mathrm{CU}}(t) \mid k\in\mathcal{K}(t)\}\subset\mathcal{O}$ and $\{\boldsymbol{o}^{(m)}_{\mathrm{TG}}(t) \mid m \in \mathcal{M}(t)\}\subset\mathcal{O}$ be their current respective locations. BSs, CUs and targets are hereby indexed respectively by $b\in\mathcal{B},k\in\mathcal{K}(t), m\in\mathcal{M}(t)$. Let us assume $\mathcal{M}(t)\cap\mathcal{K}(t)=\emptyset$. Fig. \ref{fig:sysmodel} depicts this scenario for $\mathcal{C}_k(t)=\{b_1,b_2,b_3\}$, $\mathcal{S}_m(t)=\{b_2,b_3,b_4\}$, and $b,r\in \mathcal{S}_m$. Let $a_{b,k}(t),z_{b,m}(t)\in\{0,1\},
$ such that $a_{b,k}(t)=1$ indicates BS $b$ serves CU $k$ and $z_{b,m}(t)=1$ that BS $b$ participates in sensing target $m$ in slot $t$. The sets of BSs that serve CU $k$ and target $m$ in slot $t$ are $\mathcal{C}_k(t)
=\{b\in\mathcal{B}:a_{b,k}(t)=1\}$ and $\mathcal{S}_m(t)
=\{b\in\mathcal{B}:z_{b,m}(t)=1\}$,  respectively. The baseband signal from by each BS $b$ is \footnote{Though we can reuse $s_k$ for sensing too, that would then make the sensing depend on the `active/inactive' nature of CU $k$. It would also require us to assume that the information signal to CU $k$ is known for target estimation. Therefore, we assume a dedicated random signal.}% We assume multiple BSs can simultaneously serve a CU and/or a target. 
\begin{equation}
\boldsymbol{x}_b(t)
=
\sum_{k\in\mathcal{K}(t)}
\boldsymbol{w}_{b,k}(t)s_k(t)
+
\sum_{m\in\mathcal{M}(t)}
\boldsymbol{v}_{b,m}(t)q_{b,m}(t),
\label{eq:transmit_signal}
\end{equation}
% \begin{equation}
% \boldsymbol{x}_b(t,\tau)
% =
% \sum_{k\in\mathcal{K}(t)}
% \boldsymbol{w}_{b,k}(t)s_k(\tau)
% +
% \sum_{m\in\mathcal{M}(t)}
% \boldsymbol{v}_{b,m}(t)q_{b,m}(\tau).
% \label{eq:transmit_waveform}
% \end{equation}
where $s_k(t)$ is the information for CU $k$ and $q_{b,m}(t)$ is the sampled radar signal\footnote{We assume that any specific radar signal such as a phase-modulated continuous wave (PMCW) can be adopted using a coded sequence.}. $\boldsymbol{w}_{b,k}(t)\in\mathbb{C}^{N_t}$ and $\boldsymbol{v}_{b,m}(t)\in\mathbb{C}^{N_t}$ are the respective precoders. Thus, BSs in $\mathcal{C}_k(t)$ collaboratively transmit same $s_k$ to CU $k$. We further assume that each precoder is set to zero when the particular BS is not associated with a CU/target. The communication and sensing symbols are unit power and are uncorrelated across other CUs and targets, $\mathbb{E}[s_k^*(t)s_{k'}(t)]
= \delta_{k(t)k'(t)},
\mathbb{E}[q_{b,m}(t)^*q_{b',m'}(t)]
= \delta_{bb'}\delta_{mm'},\mathbb{E}[s_k^*(t)q_{b,m}(t)]
= 0$ where $\delta_{ij}$ is Kronecker delta which takes $1$ only when $i=j$. %The sensing signals have unit power $\mathbb{E}[|q_{b,m}|^2]=1$ and uncorrelated with communication 
%symbols $\mathbb{E}[s^{*}_kq_{b,m}]=0$ for all $k,b,m$. 

% The communication waveforms have unit average power and are uncorrelated
% across CUs, $\frac{1}{\Delta}\int_0^\Delta
% \mathbb{E}\!\left[s_k^*(\tau)s_{k'}(\tau)\right]
% \,\mathrm{d}\tau
% &= \delta_{kk'}$ where $\delta_{ij}$ denotes the Kronecker delta taking $1$ only when $i=j$ and the expectation is taken over the communication symbol. Similarly, for the sensing waveforms we have, $\frac{1}{\Delta}\int_0^\Delta
% q_{b,m}^*(\tau)q_{b',m'}(\tau)
% \,\mathrm{d}\tau
% = \delta_{bb'}\delta_{mm'}$. We also assume that the sensing signal and the communication signals are uncorrelated, $\frac{1}{\Delta}\int_0^\Delta
% \mathbb{E}\!\left[s_k^*(\tau)q_{b,m}(\tau)\right]
% \,\mathrm{d}\tau
% = 0$. Also, following the radar-waveform orthogonality
% assumption in~\cite{coop2}, let the radar waveforms be orthogonal across each transmit BS--target pair over delays of interest~\cite{coop2} which we denote by $\frac{1}{\Delta}\int_{0}^{\Delta}q_{b,m}^{*}(\tau-\tau_{b,r,m})q_{b',m'}(\tau-\tau_{b'r',m'})\,\mathrm{d}\tau= 0, (b,m)\neq(b',m')$ and any receive BS $r,r^\prime$. 
%The per-BS power constraint is given by
% \begin{equation}
% \sum_{k\in\mathcal{K}(t)}
% \|\boldsymbol{w}_{b,k}(t)\|^2
% +
% \sum_{m\in\mathcal{M}(t)}
% \|\boldsymbol{v}_{b,m}(t)\|^2
% \leq P_{b_{\max}}.
% \label{eq:power_constraint}
% \end{equation}

\subsection{Cooperative Communication Model}

We assume single-antenna CUs. Let $\boldsymbol{h}_{b,k}(t)\in \mathbb{C}^{N_t}$ and $\beta_{b,k}(t)=\mathbb{E}[\|\boldsymbol{h}_{b,k}(t)\|^2]$ be the channel and path loss from BS $b$ to CU $k$.
The received signal at CU $k$ is
\begin{equation}
\begin{aligned}
y_k(t)
=
\sum_{b\in\mathcal{B}}
\boldsymbol{h}_{b,k}^{\mathrm{H}}(t)\boldsymbol{x}_b(t)
+n_k(t),
% &=\sum_{b\in\mathcal{B}}\boldsymbol{h}_{b,k}^{\mathrm{H}}(t)
% \left(\sum_{k^\prime\in\mathcal{K}(t)}
% \boldsymbol{w}_{b,k^\prime}(t)s_{k^\prime}
% +
% \sum_{m\in\mathcal{M}(t)}
% \boldsymbol{v}_{b,m}(t)q_{b,m}\right)+n_k(t),
% \\
% &=\sum_{b\in\mathcal{C}_k(t)}
% \left(\boldsymbol{h}_{b,k}^{\mathrm{H}}(t)\sum_{k\in\mathcal{K}(t)}
% \boldsymbol{w}_{b,k}(t)s_k\right)
% +\sum_{b\in\mathcal{S}_m(t)}\left(\boldsymbol{h}_{b,k}^{\mathrm{H}}(t)
% \sum_{m\in\mathcal{M}(t)}
% \boldsymbol{v}_{b,m}(t)q_{b,m}\right)+n_k(t),\\
\end{aligned}
\end{equation}%\footnote{$N_0$ is the variance, not PSD (power spectral density), as defined.}
where $n_k(t)\sim\mathcal{CN}(0,N_{0})$. Since we assumed wired backhauls and full CSI at CN, we assume coherent communication. Treating other CU and sensing signals
as interference, SINR is~\eqref{eq:sinr} at the top of the next page. The achievable rate is
\begin{figure*}[!t]
\normalsize
\begin{equation} 
\gamma_k(t) = \frac{\left| \sum_{b\in\mathcal{C}_k(t)} \boldsymbol{h}_{b,k}^{\mathrm{H}}(t)\boldsymbol{w}_{b,k}(t) \right|^2 }{ \sum_{\substack{k'\in\mathcal{K}(t)\\ k'\neq k}} \left|\sum_{b\in\mathcal{B}} \boldsymbol{h}_{b,k}^{\mathrm{H}}(t)\boldsymbol{w}_{b,k'}(t) \right|^2+ \sum_{b\in\mathcal{B}} \sum_{m\in\mathcal{M}(t)} \left|\boldsymbol{h}_{b,k}^{\mathrm{H}}(t)\boldsymbol{v}_{b,m}(t) \right|^2 +N_0 }. 
\label{eq:sinr} 
\end{equation}
\relax
\hrulefill
\end{figure*}
%Note we assume non-coherent CoMP here. We denote $\mathcal{H}_{\mathrm{CU}}=\{\tilde{\boldsymbol{h}}_{b,k}\}_{b\in\mathcal{B},k\in\mathcal{K}(t)}$.\mathbb{E}_{\mathcal{H}_{\mathrm{CU}}}[ 
\begin{equation}
R_k(t)=B_c\log_2\bigl(1+\gamma_k(t)\bigr),
\label{eq:communication_rate}
\end{equation}
for a bandwidth $B_c$. Let $Q_k(t)$, $A_k(t)$ and $D_k(t)$ respectively denote the number of pending data bits to be delivered to, number of bits arriving for, and number of bits delivered to CU $k$ in the current slot $t$. Then $D_k(t)$ satisfies
\begin{equation}
0\leq D_k(t)\leq
\min\{Q_k(t), R_k(t)\Delta\}.
\label{eq:departure_constraint}
\end{equation}
% The queue evolves according to\footnote{Since $D_k(t)\le Q_k(t)$ and $A_k(t)\ge0$, $Q_k(t+1) \ge 0$.}
% \begin{equation}
% Q_k(t+1)=Q_k(t)-D_k(t)+A_k(t).
% \label{eq:queue_evolution}
% \end{equation}
We assume CUs remain inactive ($k\notin\mathcal{K}(t)$)  as $Q_k(t)\rightarrow0$.

\subsection{Cooperative Sensing Model}
We assume each $b\in\mathcal{S}_m(t)$ jointly senses each target $m$. Targets are assumed approximately stationary during the interval of $\Delta$. For BS-target distances $d_{b,m}(t)=
\|
\boldsymbol{o}^{(m)}_{\mathrm{TG}}(t)
-
\boldsymbol{o}^{(b)}_{\mathrm{BS}}
\|,d_{r,m}(t)
=
\|
\boldsymbol{o}^{(m)}_{\mathrm{TG}}(t)
-
\boldsymbol{o}^{(r)}_{\mathrm{BS}}
\|$, the delays are $\tau_{b,r,m}(t)
=
(d_{b,m}(t)+d_{r,m}(t))/c, \quad b,r\in \mathcal{S}_m(t), m \in \mathcal{M}(t)$,
% \begin{align}
% d_{b,m}(t)
% &=
% \left\|
% \boldsymbol{o}^{(m)}_{\mathrm{TG}}(t)
% -
% \boldsymbol{o}^{(b)}_{\mathrm{BS}}
% \right\|,d_{r,m}(t)
% =
% \left\|
% \boldsymbol{o}^{(m)}_{\mathrm{TG}}(t)
% -
% \boldsymbol{o}^{(r)}_{\mathrm{BS}}
% \right\|,\\
% \tau_{b,r,m}(t)
% &=
% \frac{d_{b,m}(t)+d_{r,m}(t)}{c}, \quad b,r\in \mathcal{S}_m(t), m \in \mathcal{M}(t)
% \label{eq:sensing_geometry}
% \end{align}(received aperture at target $m$ from every BS)We assume every target reradiates its received energy isotropically.
where $c$ is the propagation speed. Let $\beta_{b,m}(t), \beta_{r,m}(t)$ denote the one-way path loss between BSs $b$,$r$ and target $m$. Let $\sigma_m(t)$ be the radar cross section (RCS) of target $m$ in square meters. Within slot $t$, let $s_k(\tau)$ denote the communication waveform from $b \in\mathcal{C}_k(t)$, and let $q_{b,m}(\tau)$ denote the sensing waveform $b\in\mathcal{S}_m(t)$ (assumed known at each receiver) where $\tau$ denotes continuous time. Thus, the waveform corresponding to~\eqref{eq:transmit_signal} is
\begin{equation}
\boldsymbol{x}_b(t,\tau)
=
\sum_{k\in\mathcal{K}(t)}
\boldsymbol{w}_{b,k}(t)s_k(\tau)
+\!\!
\sum_{m\in\mathcal{M}(t)}
\boldsymbol{v}_{b,m}(t)q_{b,m}(\tau).
\label{eq:transmit_waveform_actual}
\end{equation}%We assume it is also orthogonal to the communication waveform over delays of interest, .
Let the radar waveform be orthogonal across distinct transmit BS-target pairs over delays of interest~\cite{coop2} and with the communication waveform, $\int_{0}^{\Delta}q_{b,m}^{*}(\tau-\tau_{b,r,m})q_{b',m'}(\tau-\tau_{b'r',m'})\,\mathrm{d}\tau= 0$ for any $r,r'\in \mathcal{B}$ and $ (b,m)\neq(b',m')$; also $\int_{0}^{\Delta}q_{b,m}^{*}(\tau-\tau_{1})s_{k}(\tau-\tau_{2})\,\mathrm{d}\tau= 0$ for every $k$ and $(b,m)$. Further, let $\boldsymbol{\alpha}_{b,r,m}(t)=\sqrt{\sigma_m(t)\beta_{r,m}(t)\beta_{b,m}(t)}e^{j\phi_{b,r,m}}\boldsymbol{b}(\theta_{r,m}(t))
\boldsymbol{a}^{\mathrm{H}}(\theta_{b,m}(t))
$ where $\boldsymbol{a}(\theta_{b,m}(t))=\left[1,e^{j\pi\cos(\theta_{b,m}(t))},\dots,e^{j\pi(N_t-1)\cos(\theta_{b,m}(t))}
\right]^T$, $\boldsymbol{b}(\theta_{r,m}(t))=\left[1,\dots,e^{j\pi(N_r-1)\cos(\theta_{r,m}(t))}
\right]^T$ are the steering vectors and $\theta_{b,m}(t),\theta_{r,m}(t)$ denote the bearings relative to the horizontal axes. For $r\in\mathcal{S}_m(t)$, sensing observation is%=
%We define the channels and the corresponding path loss from BS $b$ to BS $r$ when $b\ne r$ as $\boldsymbol{H}_{br}(t)\in\mathbb{C}^{N_r\times N_t}$ and $\beta_{br}$ is the corresponding pathloss. Let $\boldsymbol{H}_{r,\mathrm{lc}}(t)$ be the self-leakage channel at each BS $r$. Let $\tau_{r,\mathrm{lc}}(t)$ be the delay of this link. $\widetilde{\tau}_{br}(t)$ denotes the direct propagation delay from BS $b$ to BS $r$. We also consider clutter elements in the environment indexed by $\ell\in\mathcal{L}(t)$~\cite{clutter}. Let $\beta^{\mathrm{cl}}_{b,\ell}(t),\beta^{\mathrm{cl}}_{r,\ell}(t)$ be the corresponding path losses, $\sigma^{\mathrm{cl}}_{\ell}(t)$ RCS and $\tau^{\mathrm{cl}}_{br,\ell}(t)$ the delays.and $\alpha^{\mathrm{cl}}_{br,\ell}(t)=\sqrt{\sigma^{\mathrm{cl}}_{\ell}(t)\beta^{\mathrm{cl}}_{b,\ell}(t)\beta^{\mathrm{cl}}_{r,\ell}(t)}e^{j\phi^{\mathrm{cl}}_{br,\ell}}\boldsymbol{b}\!\left(\theta^{\mathrm{cl}}_{r,\ell}(t)\right)\boldsymbol{a}^{\mathrm{H}}\!\left(\theta^{\mathrm{cl}}_{b,\ell}(t)\right)$ 
\begin{equation}
\begin{aligned}
\!\boldsymbol{y}_{r}\!(t,\!\tau)\!=\!\!\sum_{\!m\in\!\mathcal{M}\!(t)\!}\sum_{b\in\!\mathcal{S}_m(t)}\!\!
\!\boldsymbol{\alpha}_{br,\!m}\!(t)
\boldsymbol{x}_b(t\!,\tau-\!\tau_{b,r,m}\!(t)\!)\!\!
 +\!\boldsymbol{n}_{r\!}(t,\!\tau),
\label{eq:sensing_observation_actual}
\end{aligned}
\end{equation}
where $\boldsymbol{n}_{r}(t,\tau)\sim
\mathcal{CN}(\boldsymbol{0},N_0\boldsymbol{I}_{N_r})$. For ease of analysis, inter-BS, self, and clutter interference are assumed to be negligible by exploiting CSI and knowledge of radar signals~\cite{coop2}. %Also,  are assumed negligible after exploiting CSI and knowledge of transmitted radar signals~\cite{coop2}.

%Substituting~\eqref{eq:transmit_waveform_actual} into~\eqref{eq:sensing_observation_actual} identifies the desired sensing and communication contributions in \eqref{eq:sensing_observation_actual}. from the first term in the sensing observation Similarly, due to the orthogonality assumption for the radar waveforms across delays of interest, we are able to separate each target's echo from others.
CUs and other targets' contributions can be eliminated using the orthogonality assumptions by employing matched filtering. Let the target estimates available at the beginning of slot $t$ be $\{\hat{\boldsymbol{o}}_{\mathrm{TG}}^{(m)}(t)\mid m\in \mathcal{M}(t)\}$. We assume a unit-norm combiner at each BS $\boldsymbol{u}_{r,m}(t)
=\boldsymbol{b}(\hat{\theta}_{r,m}(t))/
\|\boldsymbol{b}(\hat{\theta}_{r,m}(t))\|$ where $\hat{\theta}_{r,m}$ is the bearing angle calculated using $\hat{\boldsymbol{o}}_{\mathrm{TG}}^{(m)}$. This combiner is fixed and not optimized. We get scalar observation
\begin{equation}
\widetilde y_{r,m}(t,\tau)
=
\sum_{b\in\mathcal{S}_m}
\widetilde{\alpha}_{b,r,m}(t) q_{b,m}(t,\tau-\tau_{b,r,m})
+
\widetilde n_{r,m}(t,\tau),
\label{eq:reduced_sensing_observation}
\end{equation}
where $\widetilde{\alpha}_{b,r,m}(t)
\!=\!\!
\boldsymbol{u}_{r,m}^{\mathrm H}
\boldsymbol{\alpha}_{b,r,m}(t)\boldsymbol{v}_{b,m}(t)$ and $\widetilde n_{r,m}(t,\tau)\sim\mathcal{CN}(0,N_0)$.  Assuming non-coherent processing, for unknowns $[\boldsymbol{o}^{(m)}_{\mathrm{TG}}(t)(0),\boldsymbol{o}^{(m)}_\mathrm{TG}(t)(1),\mathcal{R}e\{\alpha_{b,r,m}(t), \mathcal{I}m\{\alpha_{b,r,m}(t)\}]$ \cite{mimoradar}  
\begin{equation}
\boldsymbol{J}_m(t)
=
\frac{1}{\kappa}
\sum_{b,r\in\mathcal{S}_m}
|\widetilde{\alpha}_{b,r,m}(t)|^2
\boldsymbol{z}_{b,r,m}(t)\boldsymbol{z}_{b,r,m}(t)^{\mathrm{T}},
\label{eq:bf_position_fim}
\end{equation}
is the Fisher information matrix (FIM), where
$\boldsymbol{z}_{b,r,m}(t)
=[\cos\theta_{b,m}(t)+\cos\theta_{r,m}(t),
  \sin\theta_{b,m}(t)+\sin\theta_{r,m}(t)]^{\mathrm{T}}$. We observe that $\boldsymbol{z}_{b,r,m}\boldsymbol{z}^{\mathrm T}_{b,r,m} \succeq 0$ for any $\boldsymbol{z}_{b,r,m}$, so $\boldsymbol{J}_m(t)\succeq0$. The CRLB for the variance of estimation of $\boldsymbol{o}^{(m)}_{\mathrm{TG}}(t)$ after accounting unknowns is 
$\mathrm{CRLB}_m(t)=\boldsymbol{J}_m(t)^{-1}$
under $\boldsymbol{J}_m(t) \succ0$. We assume sensing waveforms have root mean square (RMS) bandwidth $B_{\mathrm{s}}$ where $\kappa
=c^2N_{0}/(8\pi^2B_{\mathrm{s}}^2N_{\mathrm{sym}})$. $N_{\mathrm{sym}}$ is the number of radar samples in interval $\Delta$. Let $p_{b,r,m}(t)=|\widetilde{\alpha}_{b,r,m}(t)|^2=|\boldsymbol{f}_{b,r,m}(t)^{\mathrm{H}}\boldsymbol{v}_{b,m}(t)|^2$ and $\boldsymbol{f}_{b,r,m}(t)=\boldsymbol{\alpha}^{\mathrm{H}}_{b,r,m}(t)\boldsymbol{u}_{r,m}$ for convenience.
% \begin{equation}
% \text{CRLB}_m(t)
% =
% \frac{\eta_{\mathrm{nc}}}
% {g_{x,m}(t)g_{y,m}(t)-h_m(t)^2}\left[\begin{matrix}g_{x,m}(t) & h_m(t)\\
% h_m(t) & g_{y,m}(t)\end{matrix}\right].
% \label{eq:localization_crlb}
% \end{equation}
% We also have the following\cite{mimoradar}
% \begin{equation}
% \begin{aligned}
% g_{x,m}(t)
% &=
% \sum_{b,r\in\mathcal{S}_m(t)}
% |\widetilde{\alpha}_{b,r,m}(t)|^2
% (\sin\theta_{b,m}(t)+\sin\theta_{r,m}(t))^2,\\
% g_{y,m}(t)
% &=
% \sum_{b,r\in\mathcal{S}_m(t)}
% |\widetilde{\alpha}_{b,r,m}(t)|^2
% (\cos\theta_{b,m}(t)+\cos\theta_{r,m}(t))^2,\\
% h_m(t)
% &=
% -\sum_{b,r\in\mathcal{S}_m(t)}
% |\widetilde{\alpha}_{b,r,m}(t)|^2
% (\cos\theta_{b,m}(t)+\cos\theta_{r,m}(t))\\
% &\qquad{}\times
% (\sin\theta_{b,m}(t)+\sin\theta_{r,m}(t)).
% \end{aligned}
% \label{eq:crlb_geometry}
% \end{equation}
\section{Problem Formulation and Convexity Discussion}
Let $\boldsymbol{A}(t)=[a_{b,k}(t)]\in\mathbb{R}^{|\mathcal{B}|\times |\mathcal{K}(t)|}$, $\boldsymbol{Z}(t)=[z_{b,m}(t)]\in\mathbb{R}^{|\mathcal{B}|\times |\mathcal{M}(t)|}$, $\boldsymbol{W}(t)=[\boldsymbol{w}_{b,k}(t)]\in\mathbb{C}^{N_t\times|\mathcal{B}|\times |\mathcal{K}(t)|}$, and $\boldsymbol{V}(t)=[\boldsymbol{V}_{m}(t)]\in\mathbb{C}^{N_t\times|\mathcal{B}|\times |\mathcal{M}(t)|}$ in which each $\boldsymbol{V}_m=[\boldsymbol{v}_{b,m}]\in\mathbb{C}^{N_t\times|\mathcal{B}|}$ and $\boldsymbol{D}(t)=[D_k(t)]\in\mathbb{R}^{|\mathcal{K}(t)|}$ respectively denote the accumulated decision variables. $\pi$ is a policy selecting
$\{\boldsymbol{A}(t),\boldsymbol{Z}(t),\boldsymbol{W}(t),
\boldsymbol{V}(t),\boldsymbol{D}(t)\}$
from the CSI, communication queues, and target information
available at the beginning of slot $t$. For the communication-sensing tradeoff in our ISAC system, we define the following objective
\begin{equation}
\begin{aligned}
&\!U(t)\!=\!
\!\rho\!
\frac{\sum_{k\in\mathcal{K}\!(t)\!}\!D_k\!(t)\!}
{\!\Delta\! R_{\mathrm{ref}}}
\!+\!
(1-\rho)\!
\!\sum_{m\in\!\mathcal{M}(t)}
\!\omega_m\!
\frac{d_{\mathrm{ref}}}
{\sqrt{\mathrm{Tr}\{\mathrm{CRLB}_m(t)\}}},
\label{eq:isac_utility}
\end{aligned}
\end{equation}
where $\mathrm{Tr}\{.\}$ is the trace. $\rho\in[0,1]$ balances rate and localization accuracy, $\omega_m>0$ denotes target priority, and $R_{\mathrm{ref}}>0$ and $d_{\mathrm{ref}}>0$ are normalization constants in bps and meters, respectively. The problem of interest is defined as
\begin{subequations}
\label{prob:dynamic_isac}
\begin{align}
\underset{\pi}{\operatorname{max}}
\text{ }&
U(t)\label{eq:objective}
\\
\operatorname{s.t.}
&\!\sum_{k\in\!\mathcal{K}(t)}
\!\|\!\boldsymbol{w}_{b,k}(t)\!\|^2
\!+\!\!\!\!\sum_{m\in\!\mathcal{M}(t)}\!\!
\!\|\!\boldsymbol{v}_{b,m}(t)\!\|^2
\!\leq \!P_{{\max}},\forall b,\label{constraint:3}\\%\nonumber\\&
&\!\min\{\!Q_k(t)\!,\!r_{k_{min}}\!\Delta\!\}\!
\!\leq \!D_k(t)\!\leq\!\min\{\!Q_k(t)\!,\!R_{k}\Delta\!\}, \forall k,\label{constraint:4}
\\
&
a_{b,k}(t),z_{b,m}(t)\in\{0,1\}, \forall b,k,m,\label{constraint:6}\\
&\|\boldsymbol{w}_{b,k}(t)\|^2
\leq a_{b,k}(t)P_{{\max}}, \forall b,k,
\label{constraint:1}\\
&\|\boldsymbol{v}_{b,m}(t)\|^2
\leq z_{b,m}(t)P_{{\max}}, \forall b,m,
\label{constraint:2}\\
&
\sum_{b\in\mathcal{B}}a_{b,k}(t)
\leq L_{\max}, \forall k,\label{constraint:7}
\\
&
\sum_{b\in\mathcal{B}}z_{b,m}(t)
\leq N_{\max}, \forall m,\label{constraint:8}\\
&
\sum_{k\in\mathcal{K}(t)}a_{b,k}(t)+\!\!\sum_{m\in\mathcal{M}(t)}\!z_{b,m}(t)
\leq C_{\max},
\quad \forall b.\label{constraint:5}
\end{align}
\label{eq:problem}
\end{subequations}
Problem \eqref{eq:problem} is a real-time optimization problem that solves for optimal BS-CU, BS-target associations, transmit beamformers, and packet departures, for a unified rate-CRLB utility objective, with a tradeoff parameter $\rho$. A per-BS power constraint is given in~\eqref{constraint:3} where $P_{{\max}}$ is the maximum power each BS can transmit at. A minimum per-CU performance is assured in~\eqref{constraint:4} where $r_{k_{\min}}$ is the threshold in bits/sec. Beamforming aligns with $(\boldsymbol{A},\boldsymbol{Z})$ according to~\eqref{constraint:1} and ~\eqref{constraint:2}. Per-CU and per-target limits on the number of BSs are placed by~\eqref{constraint:7},~\eqref{constraint:8}, while~\eqref{constraint:5} places on the total number of entities that can be connected to a BS.

Problem \eqref{eq:problem} is non-convex for three reasons: the   discrete domain of $\boldsymbol{A}$ and $\boldsymbol{Z}$, the rate $R_k(t)$, and $S(\boldsymbol{V}(t))\! =\! \sum_{m\in\mathcal{M}(t)}\! \ell_m/(\sqrt{\mathrm{Tr}\{(\boldsymbol{J}_m(t))^{-1}\}})$ where $\ell_m\!=\!((1-\rho)\omega_m d_{\mathrm{ref}})$. In the next section, we address each of these and propose an efficient low-complexity solution.

% The first one lies in the objective, which has two terms -- (i) sum of normalized departures, (ii) inverse $\sqrt{\mathrm{CRLB}_m(t)}$ with priorities. The first term is affine in $\boldsymbol{D}$. The FIM, $\boldsymbol{J}_m(t)=g(\boldsymbol{V}_m(t))$ is shown to be convex in Appendix~\ref{app:FIMconvexity}. The second term has the transformation, 
% $h(\boldsymbol{X}) = 1/\sqrt{\mathrm{Tr}\{\boldsymbol{X^{-1}}\}}$ on the FIM, $\boldsymbol{J}_m(t)\succ0$. $1/\mathrm{Tr}\{\boldsymbol{X^{-1}}\}$ can be shown to be concave on $\boldsymbol{X}\succ0$. Since square root is concave and non-decreasing on $[0,\infty)$, $1/\sqrt{\mathrm{Tr}\{\boldsymbol{X^{-1}}\}}$ remains concave when $X\succ 0$. But since $\boldsymbol{J}_m(t)$ isn't strictly affine, we cannot be sure that the composition $1/\sqrt{\mathrm{Tr}\{\boldsymbol{J}_m(t)\}}=h\circ g (\boldsymbol{V}_m(t))$ or the utility $S(\boldsymbol{V}(t)) = \sum_{m\in\mathcal{M}(t)} \ell_m/(\sqrt{\mathrm{Tr}\{(\boldsymbol{J}_m(t))^{-1}\}})$ is concave where $\ell_m=((1-\rho)\omega_m d_{\mathrm{ref}})$.

% The second source of non-convexity comes from~\eqref{constraint:4} where the $R_k(t)$ term which is the Shannon rate in terms of $\gamma_k(t)$ which in turn is a function of the beamformers. The final source of non-convexity stems from the binary set $\{0,1\}$ and hence across the association constraints.

\section{Proposed Methodology}
The proposed algorithm by definition is to be evaluated independently~$\forall t$. So, in this section, we omit the index $t$. 

\subsection{Joint Communication and Sensing Beamforming}

In this section, we first consider the continuous beamforming subproblem when we assume a specific  $(\boldsymbol{A},\boldsymbol{Z})$. 
%Also, by~\eqref{constraint:4}, $\boldsymbol{D}$ is clearly bounded by the rate term which we we improve by performing beamforming. In other words, since $\boldsymbol{W}$ and $\boldsymbol{D}$ are clearly related, it doesn't fully make sense to fix one and maximize the other alternatively. Conversely, once we maximize the objective for the variables $\boldsymbol{W},\boldsymbol{V}$, the optimal value for each $D_k = \min(Q_k,R_k\Delta)$ where $R_k$ is the improved rate after $\boldsymbol{W},\boldsymbol{V}$ are optimized.
\subsubsection{Transformed Problem} We address the first source of non-convexity of the problem -- non-affine nature of $\boldsymbol{J}_m$. We replace the quadratic gains $p_{b,r,m}$ in $\boldsymbol{J}_m$ with its bound~\cite{mm}
\begin{equation}
\begin{aligned}
|\widetilde{\alpha}_{b,r,m}|^2 &\ge 2\mathcal{R}e\left\{
\widetilde{\alpha}_{b,r,m}
(\widetilde{\alpha}_{b,r,m}^{(i)})^*\right\}-|\widetilde{\alpha}_{b,r,m}^{(i)}|^2=\widetilde{p}^{(i)}_{b,r,m},
\end{aligned}
\label{eq:bf_echo_bound}
\end{equation}
where
$\widetilde{\alpha}_{b,r,m}^{(i)}
=\boldsymbol{f}_{b,r,m}^{\mathrm{H}}
\boldsymbol{v}_{b,m}^{(i)}$.
$\widetilde{p}_{b,r,m}^{(i)}$ is affine in $\boldsymbol{v}_{b,m}$ and correspondingly
$\widetilde{\boldsymbol{J}}_m^{(i)}=\widetilde{g}(\boldsymbol{V}_m)\preceq\boldsymbol{J}_m$ is in $\boldsymbol{V}_m$. So, the composition $1/\sqrt{\mathrm{Tr}\{(\widetilde{\boldsymbol{J}}_m^{(i)})^{-1}\}}=h\circ \widetilde{g} (\boldsymbol{V}_m)$ and hence $\widetilde{S}_i(\boldsymbol{V}) = \sum_{m\in\mathcal{M}} \ell_m/(\sqrt{\mathrm{Tr}\{(\widetilde{\boldsymbol{J}}_m^{(i)})^{-1}\}})$ is concave as desired under $\widetilde{\boldsymbol{J}}_m^{(i)}\succ0$. Under $\widetilde{\boldsymbol{J}}_m^{(i)}\succ0$,  $(\widetilde{\boldsymbol{J}}_m^{(i)})^{-1}\succeq\boldsymbol{J}^{-1}_m$. Thus, by replacing the quadratic gain, we maximize a lower bound of $S(\boldsymbol{V})$. Then, for a closed-form solution, we apply the Lipschitz descent lemma~\cite{boyd2004convex} for $\Lambda>0$ obtained by backtracking
\begin{equation}
\resizebox{1\linewidth}{!}{%
$\begin{aligned}
\widetilde{S}_i(\boldsymbol{V}) \ge& \widetilde{S}_i(\boldsymbol{V}^{(i)}) + 2\mathcal{R}e\Bigg\{\sum_{b,m} \left( \!\nabla_{\!\boldsymbol{v}_{b,m}^*}\!\!\! \left.\widetilde{S}_i\right|_{\!\boldsymbol{V}^{\!(i)\!}} \right)^{\mathrm{H}} (\boldsymbol{v}_{b,m}-\boldsymbol{v}_{b,m}^{(i)})\Bigg\}  \\
& - \frac{\Lambda}{2} \sum_{b,m} \|\boldsymbol{v}_{b,m} - \boldsymbol{v}_{b,m}^{(i)}\|^2=\underline{S}_i(\boldsymbol{V}),
\end{aligned}$%
}
\end{equation}

where $\boldsymbol{V}^{(i)}$ is the previous update and  
\begin{equation}
\begin{aligned}
\!\nabla_{\!\boldsymbol{v}_{b,m}^*}\!\!\! \left.\widetilde{S}_i\right|_{\!\boldsymbol{V}^{\!(i)\!}}\!\!&=\!\!\Bigg\{\!\sum_{\!r\!\in\mathcal{S}_m\!}\!\left(\! \frac{\!\ell_m \|(\boldsymbol{J}_m)^{\!-1\!}\boldsymbol{z}_{\!b,r,m}\!\|^2}{\!2\kappa\big[\!\mathrm{Tr}\!\big(\!(\boldsymbol{J}_m)^{\!-1}\!\big)\!\big]^{\!1.5}}\!\!  \right)\!\!\boldsymbol{f}_{b,r,m}\boldsymbol{f}_{b,r,m\!}^{\!\mathrm{H}}\!\Bigg\}\! \boldsymbol{v}_{\!b,m\!}^{(i)}. 
\end{aligned}
\end{equation}
The new surrogate $\underline{S}_i(\boldsymbol{V})\leq\widetilde{S}_i(\boldsymbol{V})\le S(\boldsymbol{V})$ remains concave. We then apply fractional programming to remove the logarithm and decouple the $\gamma_k$ ratio in $R_k$.  $R_k$ after applying Lagrangian dual transform (LDT) is~\cite{fp2}
\begin{equation}
\begin{aligned}
\widetilde{R}^{\!(i)}_k\!(\!\boldsymbol{W}\!,\!\boldsymbol{V}\!)\!=\!  \frac{B_c}{\ln{2}}\!\Big[ \!\ln(1\!+\!\eta^{\!(i)}_k\!)\! -\! \eta^{\!(i)}_k\! +\! (1+\eta^{\!(i)}_k)\!\frac{\gamma_k}{1+\gamma_k}\Big].
\end{aligned}
\end{equation}
As long as $\eta^{(i)}_k=\gamma_k$, $\widetilde{R}^{(i)}_k=R_k$. Thus, the auxiliary update is $\eta^{(i)}_k=\gamma^{(i)}_k$ where $\gamma^{(i)}_k=\gamma_k(\boldsymbol{W}^{(i)},\boldsymbol{V}^{(i)})$. Then, let $\mu_k(\boldsymbol{W})
=\sum_{b\in\mathcal{B}}
\boldsymbol{h}_{b,k}^{\mathrm{H}}\boldsymbol{w}_{b,k}$ and let
$I_k(\boldsymbol{W},\boldsymbol{V})$ denote the denominator in~\eqref{eq:sinr} (noise-plus-interference power). We then apply quadratic transform (QT) to $\widetilde{R}^{(i)}_k$~\cite{fp1}
\begin{equation}
\begin{aligned}
&\!\underline{R}_k^{(i)}(\boldsymbol{W},\!\boldsymbol{V})
\!=\!{}\frac{B_c}{\ln 2}\!\Bigg[\! \ln(1+\eta^{(i)}_k)\! - \eta^{(i)}_k\! +\left(\! 2\sqrt{\!1+\eta^{\!(i)\!}_k\!}\right)\\ &\times\left(\mathcal{R}e\!\Big\{\! (\!\zeta^{\!(i)}_k\!)^*\!\mu_k\!(\!\boldsymbol{W}\!)\! \Big\}\right) - |\zeta^{(i)}_k|^2\! \Big(\!I_k(\boldsymbol{W}\!,\!\boldsymbol{V})\! +\! |\mu_k(\!\boldsymbol{W}\!)|^2\Big)\Bigg],
\end{aligned}
\label{eq:bf_rate_bound}
\end{equation}
where the auxiliary variable is updated using
\begin{equation}
\zeta_k^{(i)}
=
\frac{\sqrt{(1+\eta_k^{(i)})}\mu_k(\boldsymbol{W}^{(i)})}
{I_k(\boldsymbol{W}^{(i)},\boldsymbol{V}^{(i)})+|\mu_k(\boldsymbol{W}^{(i)})|^2}.
\label{eq:bf_fp_auxiliary}
\end{equation}
Since $\underline{R}_k^{(i)}$ is concave in $(\boldsymbol{W},\boldsymbol{V})$, the constraint~\eqref{constraint:4} is convex. Having fixed the association variables, the transformed continuous beamforming subproblem is convex. 

\subsubsection{Beamforming Design} By fixing $\boldsymbol{A},\boldsymbol{Z}$, we set $\boldsymbol{w}_{b,k}=0,\text{ } \forall b \notin \mathcal{C}_k,\text{ } \boldsymbol{v}_{b,m}=0 \text{ }\forall b  \notin \mathcal{S}_m$ making constraints~\eqref{constraint:1},~\eqref{constraint:2} redundant. Constraints~\eqref{constraint:6}-\eqref{constraint:5} are irrelevant too. We set $\chi = \{(\boldsymbol{W},\boldsymbol{V},\boldsymbol{D}) \mid \min{\{Q_k,r_{k_{\min}}}\Delta\}\le D_k \le Q_k \text{ }\forall k, \underline{\boldsymbol{J}}^{(i)}_m \succ0 \text{ }\forall m\}$ as the domain of the primal problem. Let $\boldsymbol{\lambda} =\{\lambda_b\ge0, \forall b\}$ and $\boldsymbol{\nu}=\{\nu_k\ge0, \forall k\}$ be the dual variables for~\eqref{constraint:3} and $D_k\leq R_k\Delta$ respectively. The Lagrangian for the $i^{\mathrm{th}}$ iteration of the transformed problem is
\begin{equation}
\resizebox{0.85\linewidth}{!}{%
$\begin{aligned}
        \mathcal{L}^{(i)}&(\boldsymbol{W},\boldsymbol{V},\boldsymbol{D},\boldsymbol{\lambda},\boldsymbol{\nu})= 
\frac{\rho\sum_{k\in\mathcal{K}}D_k}
{\Delta R_{\mathrm{ref}}}
+
\underline{S}_i(\boldsymbol{V}) \\
&-\sum_{b \in \mathcal{B}}\lambda_b\left(\sum_{k\in\mathcal{K}}
\|\boldsymbol{w}_{b,k}\|^2
+ \sum_{m\in\mathcal{M}}
\|\boldsymbol{v}_{b,m}\|^2
- P_{{\max}}\right)\\
&-\sum_{k \in \mathcal{K}}\nu_k\left(D_k - \underline{R}^{(i)}_k(\boldsymbol{W},\boldsymbol{V})\Delta\right).
\end{aligned}$}
\end{equation}
Let $\widetilde{\boldsymbol{w}}_k = [\boldsymbol{w}_{b,k}]_{b \in \mathcal{C}_k} \in \mathbb{C}^{|\mathcal{C}_k|N_t\times1}$, $\widetilde{\boldsymbol{h}}_{\mathcal{C}_k,k'}=[\boldsymbol{h}_{b,k'}]_{b \in \mathcal{C}_k}\in \mathbb{C}^{|\mathcal{C}_k|N_t\times1}$ be the be the accumulated beamformers to CU $k$ and channel matrices from each BS in $\mathcal{C}_k$ to an arbitrary CU $k'$, respectively. $\mathcal{L}^{(i)}$ is rewritten accordingly and a regularization $-\epsilon\sum_{k\in\mathcal{K}}\|\widetilde{\boldsymbol{w}}_k-\widetilde{\boldsymbol{w}}^{(i)}_k\|^2$ is added to ensure we get an update with an invertible matrix. Let
$\boldsymbol{\Lambda}_k = \mathrm{blkdiag}\left([(\lambda_b+\epsilon) \boldsymbol{I}_{N_t}]_{b\in\mathcal{C}_k}\right)\in \mathbb{R}^{|\mathcal{C}_k|N_t\times|\mathcal{C}_k|N_t}$ for a small $\!\epsilon\!>\!0$. Thus the stationary conditions using the Lagrangian with respect to each
$\widetilde{\boldsymbol{w}}_k$ yields the accumulated update
\begin{equation}
\begin{aligned}
&\widetilde{\boldsymbol{w}}^{(i+1)}_k\!=\!\left(\! \boldsymbol{\Lambda}_k\!+\! \sum_{\!k'\!\in\mathcal{K}} \!d^{(i)}_{k'} \widetilde{\boldsymbol{h}}_{\!\mathcal{C}_k\!,k'} \widetilde{\boldsymbol{h}}_{\mathcal{C}_k,k'}^{\mathrm{H}}\! \right)^{\!\!-1}\!\!\!
\left(\!c_k^{(i)}\widetilde{\boldsymbol{h}}_{\mathcal{C}_k,\!k}\!+\!\epsilon\widetilde{\boldsymbol{w}}_k^{\!(i)}\!\right),
\end{aligned}
\label{eq:w_update}
\end{equation}
where $\!c_k^{(i)}\! \!\!=\! (\zeta^{(i)}_k\!\nu_{k} \Delta \!B_c \!\sqrt{1\!+\!\eta^{(i)}_k})/\!\ln \!2,d_{k'}^{(i)}\! \!\!=\! (\!|\zeta^{(i)}_{k'}\!|^2\!\nu_{k'}\! \Delta \!B_c)\!/\!\ln \!2$. This form is much more tractable than an $\boldsymbol{w}_{b,k}$ update. %The accumulated update is also meaningful for the coherent communication model.

Similarly, using stationary conditions for $\boldsymbol{v}_{b,m}$, we get
% \begin{equation}
% \begin{aligned}
% &\boldsymbol{v}_{\!b,m\!}^{\!(i+1)}\!\!\!= \!\!\!\Bigg(\!\!\!\left(\!\! \lambda_b\!\!+\!\! \!\frac{\Lambda}{2}\! \!\right)\!\!\boldsymbol{I}_{\!N_t}\!\!\!+\!\!\! \sum_{\!k'\in\!\mathcal{K}\!}\! d_{k^\prime}^{(i)} \!\boldsymbol{h}_{\!b,k'}\! \boldsymbol{h}_{\!b,k'}^{\mathrm{H}}\!\!\Bigg)^{\!\!\!-1}\!\!\!\! \!\left(\! \!\nabla_{\!\boldsymbol{\!v\!}_{\!b,m\!\!}^*\!}\!\! \widetilde{S}_i(\!\boldsymbol{V}^{\!(i)\!})\!\! +\!\! \frac{\Lambda}{2}\!\!\boldsymbol{v}_{\!b,m}^{\!(i)}\!\!\! \right)\!\!.
% \end{aligned}
% \label{eq:v_update}
% \end{equation}
\begin{equation}
\begin{aligned}
&\!\!\boldsymbol{v}_{\!b,m\!}^{\!(i+1)}\!\!\!= \!\!\Bigg(\!\!\left(\!\! \lambda_b\!\!+\! \!\frac{\Lambda}{2} \!\right)\!\!\boldsymbol{I}_{\!N_t}\!+\!\! \sum_{\!k'\in\!\mathcal{K}\!}\! d_{k'}^{(i)}\boldsymbol{h}_{\mathrm{eff}_{b,k'}} \!\!\Bigg)^{\!\!\!-1}\!\!\!\! \!\left(\! \!\nabla_{\boldsymbol{\!v\!}_{\!b,m\!\!}^*\!}\!\widetilde{S}_i(\!\boldsymbol{V}^{\!(i)\!})\! +\! \frac{\Lambda}{2}\boldsymbol{v}_{\!b,m}^{\!(i)}\!\! \right)\!\!,
\end{aligned}
\label{eq:v_update}
\end{equation}
where $\boldsymbol{h}_{\mathrm{eff}_{b,k'}}\!=\!\boldsymbol{h}_{b,k'}\boldsymbol{h}_{b,k'}^{\mathrm{H}}$. Since $\Lambda>0$, the matrix is invertible. Overall procedure is outlined in Algorithm~\ref{alg:beamforming}. $\epsilon,\epsilon_{\lambda},\epsilon_\nu$ represent the tolerances for $U(t)$ and $(\boldsymbol{\lambda},\boldsymbol{\nu})$. %At $\boldsymbol{V}^{(i)}$, since the surrogates are tight, it converges optimally for sensing. 
% In every iteration $i$,  $\eta_k^{(i)}=\gamma^{(i)}_k$ and \eqref{eq:bf_fp_auxiliary} are updated. For each $\Lambda$,~\eqref{eq:w_update} and \eqref{eq:v_update} are alternated with bisection searches for $\lambda_b$ and $\nu_k$ until~\eqref{constraint:3} and~\eqref{constraint:4} hold with complementary slackness. Each of these updates are repeated for each $\Lambda\rightarrow\mu \Lambda$ for some $\mu>1, \Lambda_{\mathrm{init}}>0$ until the assumed $\underline{\boldsymbol{J}}_m^{(i)}(\boldsymbol{V})\succ0,\ \forall m$, and $\widetilde{S}_i(\boldsymbol{V})\ge\underline{S}_i(\boldsymbol{V})$ satisfies. Departures are then set to $D_k=\min\{Q_k,R_k^{(i)}\Delta\}$. The process is repeated until convergence to solve the surrogate beamforming problem.
\setlength{\textfloatsep}{5pt}
\begin{algorithm}[t]
%\vspace{0.06in}
\caption{Beamforming Algorithm}
\label{alg:beamforming}
\small
\begin{algorithmic}[1]
\State Initialize $\boldsymbol{W}^{(0)},\boldsymbol{V}^{(0)}$, $\Lambda_{\mathrm{init}}>0$, $\mu>1$, and $i=0$.
\Repeat
    \State Update $\eta_k^{(i)}=\gamma_k^{(i)}$ and \eqref{eq:bf_fp_auxiliary}; set $\Lambda=\Lambda_{\mathrm{init}}/\mu$.
    \Repeat
        \State Set $\Lambda\leftarrow\mu\Lambda$.
        \State Alternate \eqref{eq:w_update} and \eqref{eq:v_update}, bisecting each $\lambda_b$ until \eqref{constraint:3} holds and setting $\nu_k=\rho/(\Delta R_{\mathrm{ref}})$, bisected below it until $R_k^{(i)}\Delta=Q_k$ or above it until $R_k^{(i)}\Delta=\min\{Q_k,r_{k\min}\Delta\}$
        % \State Alternate \eqref{eq:w_update} and \eqref{eq:v_update} with $\nu_k = \rho / (\Delta R_{\mathrm{ref}})$ for $Q_k>R_{\mathrm{ref}}\Delta$ and otherwise bisected either side until \eqref{constraint:4} holds with complementary slackness and $\lambda_b$ until \eqref{constraint:3}
    \Until{$\underline{\boldsymbol{J}}_m^{(i)}(\boldsymbol{V})\succ \boldsymbol{0},\ \forall m$, and $\widetilde{S}_i(\boldsymbol{V})\ge\underline{S}_i(\boldsymbol{V})$}
    % \State Update $\boldsymbol{W}^{(i+1)}\leftarrow\boldsymbol{W},\ \boldsymbol{V}^{(i+1)}\leftarrow\boldsymbol{V}$.
    \State Set $D_k=\min\{Q_k,R_k^{(i)}\Delta\},\ \forall k$; $i\leftarrow i+1$.
\Until{convergence with tolerances $\epsilon,\epsilon_{\lambda},\epsilon_{\nu}$}
\end{algorithmic}
\end{algorithm}
\setlength{\textfloatsep}{12pt}
\subsection{Association Strategy and the Overall Solution}
We are required to isolate $(\boldsymbol{A},\boldsymbol{Z})$ from the beamforming problem, due to their non-convex domain. Fixed distance-based association (FA) where each CU and target are mapped to their nearest BSs subject to~\eqref{constraint:7},~\eqref{constraint:8},~\eqref{constraint:5} is noted optimal for pathloss dependent CUs, though there are other geometric and random factors to be considered for sensing~\cite{coop1}. 

Therefore, we propose a dynamic association (DA) which also assumes $\boldsymbol{A}, \boldsymbol{Z}$ are fixed by and initialized by distance, respectively. An exhaustive search (before considering~\eqref{constraint:5}) involves ${|\mathcal{B}| \choose N_{\max}}^{|\mathcal{M}|}$ choices for $\boldsymbol{Z}$ -- making it impractical for a real-time algorithm. Therefore, we propose to restrict the search to one change in $\boldsymbol{Z}$ (i.e., one target-BS association to be added, deleted or updated) after each beamforming update using  $S(\boldsymbol{V}^{(i)})$.  Updates that do not obey~\eqref{constraint:5} are rejected. If a new BS is mapped to a target, then we use a model $\boldsymbol{v}^{(i)}_{b,m}=(P_{\max}\boldsymbol{a}^{\mathrm{H}}(\hat{\theta}_{b,m}))/|\mathcal{M}||\mathcal{K}||\boldsymbol{a}^{\mathrm{H}}(\hat{\theta}_{b,m})|$ subject to ~\eqref{constraint:3}. We repeat this strategy until convergence.
% $\underline{R}_k^{(i)}\Delta$ is $>Q_k$ or $<\min\{Q_k,r_{k_{\min}}\Delta\}$, in which case $\nu_k$ is found by bisection

\begin{table}[t]
\caption{Per-outer-iteration complexity Analysis.}
%\vspace*{0.04in}
\label{tab:beamforming_complexity}
\centering
\renewcommand{\arraystretch}{1.4}
\setlength{\extrarowheight}{2pt}
\resizebox{\columnwidth}{!}{%
\begin{tabular}{|l|c|}
\hline
\textbf{Considered Problem} & \textbf{Complexity} \\
\hline
$(\boldsymbol{W},\boldsymbol{V})$ updates
&
\multirow{2}{*}{$\mathrm{O}\left(C_{\mathrm{bf}}=
|\mathcal B|\left(\left(|\mathcal C_k|N_t\right)^3
+N_t^3\right)\right)$}
\\
for fixed $\boldsymbol{\lambda},\boldsymbol{\nu},\Lambda$ & \\
\hline
Beamforming Subproblem/
&
\multirow{2}{*}{$\mathrm{O}\!\left(\!
N_{\mathrm{bt}}N_{\mathrm{dual}}
(|\mathcal{B}|N_{\lambda}+|\mathcal{K}|N_{\nu})
C_{\mathrm{bf}}\right)$}
\\
Overall Problem&\\
\hline
SDR based beamforming\cite{beam3}
&
$\mathrm{O}\!\left(
|\mathcal{K}|^{6.5}N_t^{6.5}
\log\!\left(1/\epsilon\right)
\right)
$
\\
\hline
\end{tabular}%
}
\end{table}
% For fixed multipliers and $L$, the complexity is as listed in Table~\ref{tab:beamforming_complexity} since inverted matrix in~\eqref{eq:v_update} remains the same for all $m \in \mathcal{M}$ served by a BS $b\in \mathcal{B}$. Search for each $\lambda_b$ and $\nu_k$ requires at most $N_\lambda=\lceil\log_2(R_\lambda/\epsilon_\lambda)\rceil$ and $N_\nu=\lceil\log_2(R_\nu/\epsilon_\nu)\rceil$ bisection steps, respectively, for bracket widths $R_\lambda,R_\nu$ and tolerances $\epsilon_\lambda,\epsilon_\nu$. Thus, the complexity for the beamforming subproblem is as listed where $N_{\mathrm{bt}}$, $N_{\mathrm{dual}}$, $N_{\mathrm{it}}$ represent the number of trials for backtracking, dual variables and for overall convergence. This is the dominant term in the complexity for the overall algorithm as well wherein $N_{\mathrm{it}}$ is larger for dynamic association (more iterations for additional convergence of $\boldsymbol{Z}$). Since the proposed method have closed--form updates coupled only with scalar variables which are computed by bisection, we avoid the need for high--complexity SDP solvers in each iteration like~\cite{beam1,beam2,beam3} or inner--loop gradient descents, we attain a much better complexity.

Then, Table~\ref{tab:beamforming_complexity} presents the complexity analysis of the overall problem. For fixed $\boldsymbol{\lambda},\boldsymbol{\nu},\Lambda$, the matrix inverse in~\eqref{eq:v_update} is shared across all $m \in \mathcal{M}$ served by a BS $b\in \mathcal{B}$. Each $\lambda_b$ and $\nu_k$ needs at most $N_\lambda=\lceil\log_2(R_\lambda/\epsilon_\lambda)\rceil$ and $N_\nu=\lceil\log_2(R_\nu/\epsilon_\nu)\rceil$ steps for widths $R_\lambda,R_\nu$. $N_{\mathrm{bt}}$ and $N_{\mathrm{dual}}$ denote the backtracking and dual iterations. This beamforming step also dominates the overall algorithm's complexity, wherein DA could take more outer iterations for $\boldsymbol{Z}$ convergence. Since our closed-form updates are coupled only through bisection-computed scalars, we avoid per-iteration SDP solvers~\cite{beam1,beam2,beam3} or inner-loop gradient descents, yielding lower complexity. %Note that \cite{beam3} denotes $\epsilon$ as the solution tolerance.
% We consider a $200\times 200\text{ m}^2$ area (Fig.~\ref{fig:setup}) with $|\mathcal{B}(t)|=10$, $|\mathcal{K}(t)|=4$, $|\mathcal{M}(t)|=6$, and $N_t=N_r=16$ per BS. Links $i\in\{(b,k),(b,m),(r,m)\}$ follow the $5$~GHz 3GPP UMi pathloss model~\cite{3gpp} with distance $d_i$ and loss $\beta_{i_{\mathrm{dB}}}$ (yielding $\beta_{i_{\mathrm{dB}}}=77.57$~dB at $d_i=10$~m with one-way exponent $3.67$, departing from the $2\text{--}2.2$ exponent in~\cite{coop2,close1,close2}). Unless varied, default parameters are $B_c=B_s=10$~MHz, $P_{\max}=40$~dBm, $N_{0_{\mathrm{dB}}}=-100$~dBm, $C_{\max}=6$, $L_{\max}=2$, $N_{\max}=4$, $r_{\min}=1$~Mbps, $Q_0(t)=\{300,500,100,200\}$~Mbits, $N_{\mathrm{sym}}=1024$~\cite{close2}, $\omega_m=1/|\mathcal{M}(t)|$, and tolerances $\epsilon=10^{-4}$, $\epsilon_\lambda=\epsilon_\nu=10^{-3}$. Setting $R_{\mathrm{ref}}=100$~Gbps, $d_{\mathrm{ref}}=1$~m, and $\Delta=1$~s balances the communication and sensing terms in $U(t)$ to the same order.
\section{Results and Discussion}
\begin{figure}[t!]
\centering
     % \begin{subfigure}[b]{0.67\columnwidth}
        \includegraphics[width=0.59
        \linewidth]{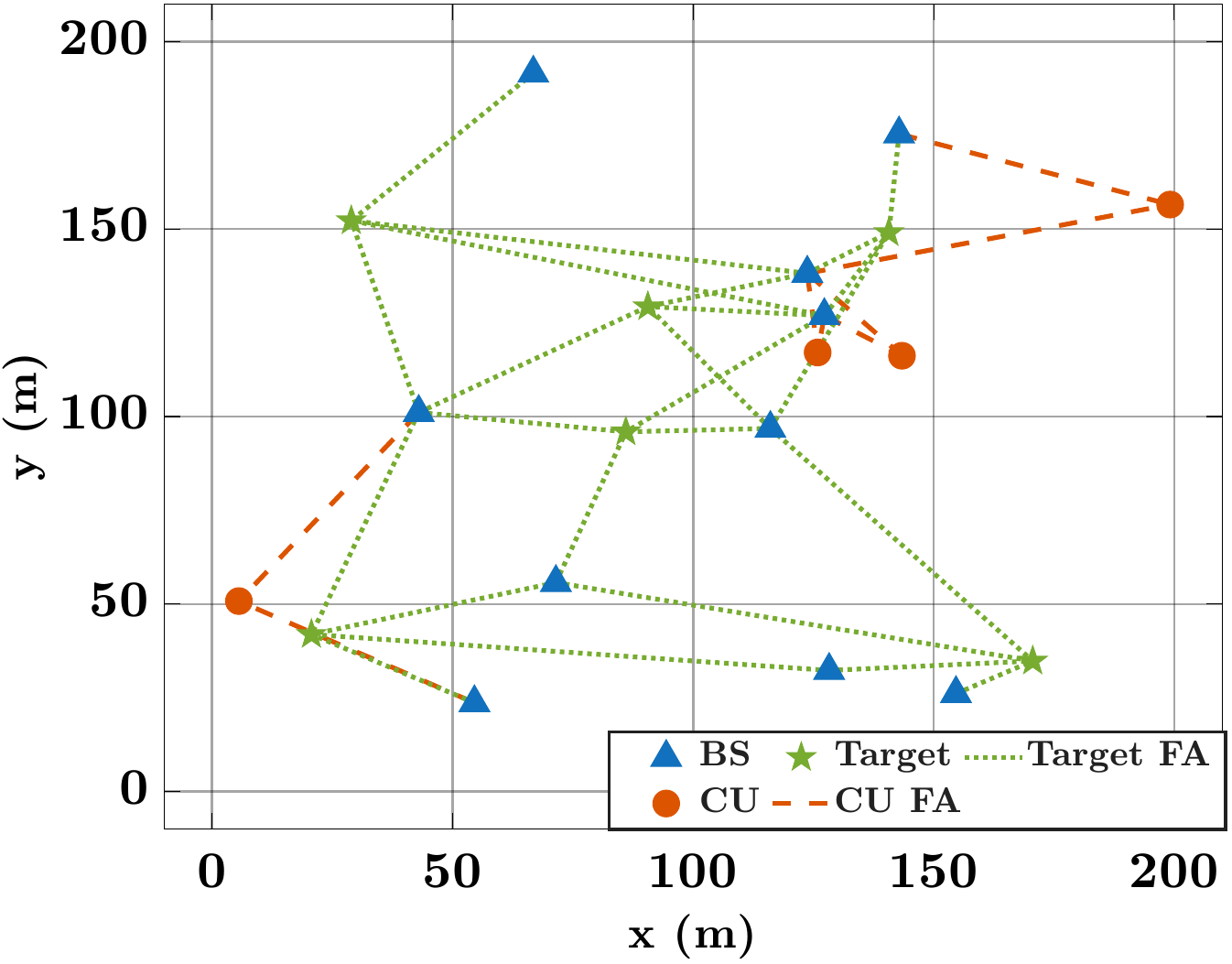}
    \vspace{-5 pt}
    \caption{Simulation setup for a $200\times200$ service area.\label{fig:setup}}
            \vspace{-8pt}
  \end{figure}

% The 3GPP urban microcell pathloss model~\cite{3gpp} is assumed. Let $d_i, \beta_{i_{\mathrm{dB}}}$ represent the distance (in meters) and pathloss (in dB) of each link . When carrier frequency is $5$ GHz, for instance, we get $\beta_{{b,k}_{\text{dB}}}=\beta_{{b,m}_{\text{dB}}}=\beta_{{r,m}_{\text{dB}}} = 77.57$ dB for $d_{b,k} = d_{b,m} = d_{r,m} =  10$~m (one-way pathloss exponents $3.67$). This is a departure from~\cite{coop2,close1,close2} which assume an exponent of $2-2.2$. We set $B_c=B_s=10$ MHz, $P_{\max}=40$ dBm, $N_{0_{\text{dB}}}= -100$ dBm. Tolerances are set as $\epsilon=10^{-4},\epsilon_\lambda=\epsilon_\nu=10^{-3}$. 
% By default, we assume $C_{\max} = 6$, $L_{\max} = 2$, $N_{\max} = 4$, $r_{\min}=1$ Mbps. We assume a queue $Q_0(t)=\{300,500,100,200\}$ in Mbits for each CU, by default. We also assume $N_{\mathrm{sym}}=1024$~\cite{close2} and $\omega_m=1/|\mathcal{M}(t)|~\forall m$.
We consider a simulation setup with $|\mathcal{B}(t)| = 10$, $|\mathcal{K}(t)| = 4$, $|\mathcal{M}(t)| = 6$, and $N_t = 16$ and $N_r = 16$ per BS as in Fig.~\ref{fig:setup}.  All links $i\in\{(b,k),(b,m),(r,m)\}$ follow the $5$~GHz 3GPP UMi pathloss model~\cite{3gpp} with distance $d_i$ and loss $\beta_{i_{\mathrm{dB}}}$ (yielding $\beta_{i_{\mathrm{dB}}}=77.57$~dB at $d_i=10$~m with one-way exponent $3.67$, different from $2-2.2$ in~\cite{coop2,close1,close2}). The default parameters are $B_c=B_s=10$~MHz, $P_{\max}=40$~dBm, $N_{0_{\mathrm{dB}}}=-100$~dBm, $C_{\max}=6$, $L_{\max}=2$, $N_{\max}=4$, $r_{\min}=1$~Mbps, $Q(t)=Q_0(t)=\{10,50,50,10\}$~kbits, $N_{\mathrm{sym}}=1024$, $\omega_m=1/|\mathcal{M}(t)|,\forall m$, and tolerances $\epsilon=10^{-4}$, $\epsilon_\lambda=\epsilon_\nu=10^{-3}$. We set $R_{\mathrm{ref}} = 100$~Gbps, $d_{\mathrm{ref}} = 1$~m, and $\Delta = 0.1$ ms to set the sensing part of $U(t)$ roughly two orders of magnitude over communication, with emphasis on sensing. Also, notably the minimum number of BSs assigned to each target is set as $2$ to avoid $\boldsymbol{J}_m\nsucc0$. 
  
For metrics, we consider \textit{average per-user rate}, $(1/|\mathcal{K}(t)|\Delta B_c)\sum_kD_k(t))$, in bps/Hz, and the \textit{harmonic mean (HM) of root CRLB (RCRLB)}, $((1/|\mathcal{M}(t)|)\sum_m(\sqrt{\mathrm{Tr}\{\boldsymbol{J}_m^{-1}\}})^{-1})^{-1}$, in meters. Notably, HM$\le$AM (arithmetic mean) since $\sqrt{\mathrm{Tr}\{\boldsymbol{J}_m^{-1}\}}>0$.

We then assume perfect sensing (PS) case denotes the availability of $\{\boldsymbol{o}^{(m)}_{\mathrm{TG}}(t)\}$ and the imperfect sensing (IS) is the case when not. To study the effect of sensing error on the proposed method, we assume a Gaussian $\{\boldsymbol{e}_m=\hat{\boldsymbol{o}}_{\mathrm{TG}}^{(m)}-\boldsymbol{o}^{(m)}_{\mathrm{TG}}\mid m\in \mathcal{M}\}\sim\mathcal{N}(\boldsymbol{0},\sigma_e^2\boldsymbol{I}_2)$.  Our implementation of~\cite{coop2} -- PS, FA case with zero forcing (ZF) beamforming\footnote{ZF is executed for $10$ iterations. Since we consider perfect CSI (for communication) and PS, with more than $>50$ iterations, it converges to the derived weighted minimum mean square error type updates.} is considered. 

\begin{figure}[t!]
\centering        \includegraphics[width=0.61
        \linewidth]{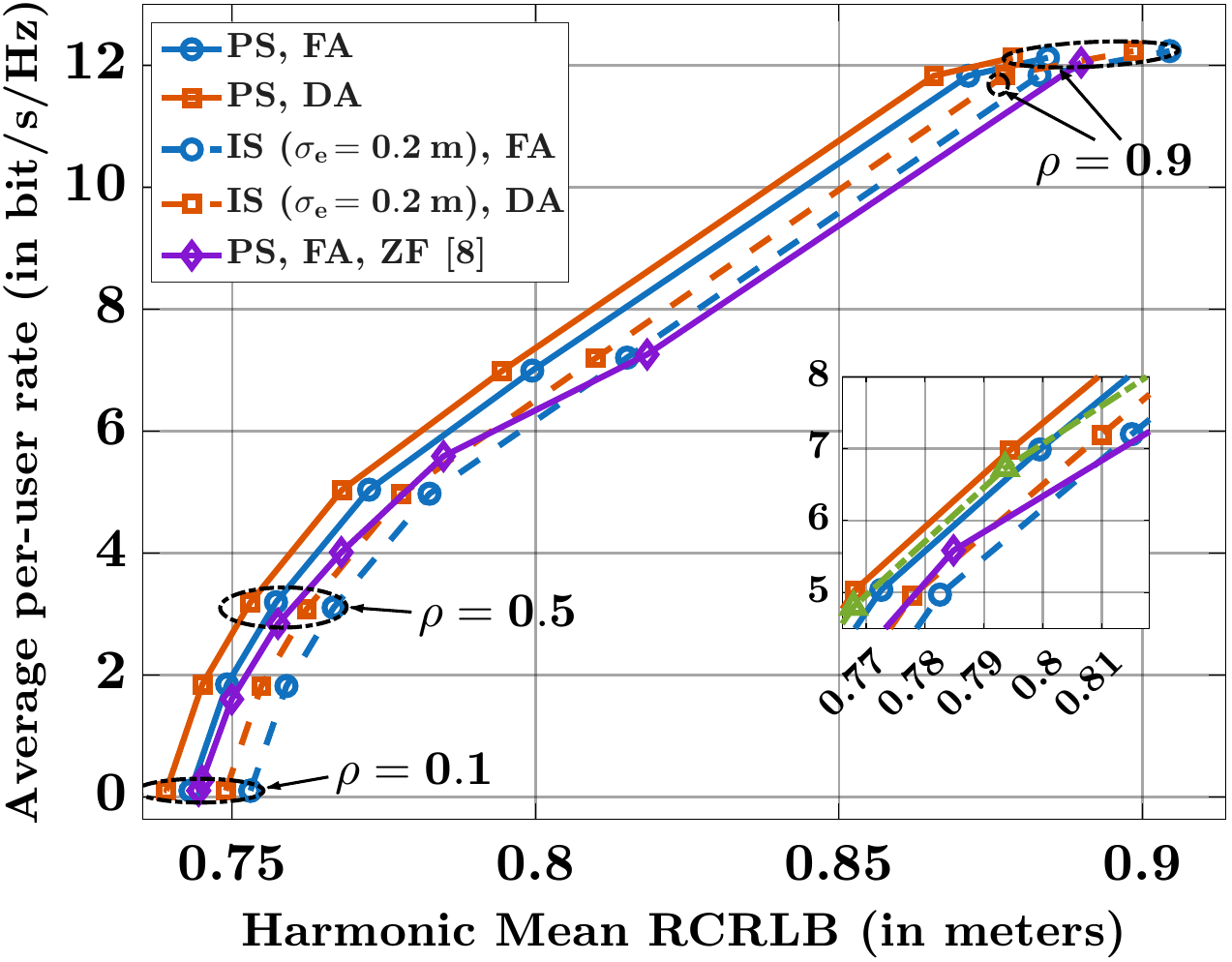}
        \vspace{-3 pt}
\caption{Comparison of Proposed DA with FA and baselines under PS/IS.\label{fig:rho_sweep}}
        \vspace{-8pt}
  \end{figure} 
% \subsection{Communication-Sensing Tradeoff and Performance Bounds}
% Due to our emphasis on sensing ($R_{\mathrm{ref}}$, $d_{\mathrm{ref}}$), in Fig.~\ref{fig:rho_sweep}, we see that in $\rho\in[0.1,0.5]$ region, the curve has a greater slope than in $\rho\in(0.5,0.9]$. At about $\rho=0.7$, we get the most fair performance of about $0.8$ m RCRLB and $7$ bps/Hz rate. For our queue-limited setup, once the queued information is delivered to all users (beyond $\rho=0.8)$, the trend saturates. The proposed DA outperforms FA consistently in sensing. This gap is wider under IS because DA tries to compensate for the target uncertainty with better sensing associations. We observe that under IS, there is a $0.01-0.15$ m decrease in RCRLB for a $\sigma_e=0.2$ m error in $\{\boldsymbol{o}^{(m)}_{\mathrm{TG}}\}$ (only 5\%), illustrating the algorithm's superiority. Notably, for $\rho\ge0.7$, rate under IS exceeds PS by $0.1-0.2$ bps/Hz -- as sensing gain saturates sooner under IS, so communication gains in this case. Finally, though the $L_{\max}=1$ baseline~\cite{close2} with the same $C_{\max}=6$ performs slightly better in sensing since more BSs are allocated for sensing, achieves worse communication performance -- a maximum gap of $3$ bps/Hz at $\rho=0.8$. The ZF~\cite{coop2} baseline achieves results similar to the IS cases. Interestingly, at around $5-7$ bps/Hz, IS, DA beats this baseline, indicating the robustness of the proposed scheme under IS.
Due to our sensing emphasis ($R_{\mathrm{ref}}, d_{\mathrm{ref}}$), Fig.~\ref{fig:rho_sweep} exhibits a steeper slope for $\rho\in[0.1,0.5]$ than for $\rho\in(0.5,0.9]$, achieving a balanced trade-off at $\rho=0.7$ ($0.8$~m RCRLB, $7$~bps/Hz). Beyond $\rho=0.8$, the rate saturates as some CU queue(s) are depleted. Across all cases, the proposed DA outperforms FA in sensing (with similar rate), with an even wider margin under IS as DA dynamically selects better associations to compensate for target uncertainty. Even with a $\sigma_e=0.2$~m error in $\{\boldsymbol{o}^{(m)}_{\mathrm{TG}}\}$, the RCRLB degrades by only $0.015$~m, confirming the algorithm's robustness. Notably, for $\rho\ge 0.7$, IS yields a $0.1$~bps/Hz higher rate than PS because sensing gains saturate earlier under uncertainty, freeing resources for communication. The ZF baseline~\cite{coop2} performs like the proposed scheme under IS and is even outperformed by IS-DA at $5\text{--}7$~bps/Hz. Despite severe two-way pathloss, similar sensing and communication metrics are observed as in~\cite{coop2,close1,close2}.

% Let us first consider Fig.~\ref{fig:rho_sweep} for $\rho\in[0,1]$. For this setup, we obtain Fig.~\ref{fig:rho_sweep} by varying $\rho\in[0,1]$.  Firstly, we observe that DA performs consistently better than FA in sensing as we update $\boldsymbol{Z}$ to improve the same. Notably, this gap widens as $\rho$ decreases (lower-right region). We also observe that an estimation error of $\sigma_e=0.2$ m introduce a noticeable degradation in sensing. Since~\cite{close2} restricts $L_{\max}=1$, it performs significantly worse than the proposed method (PS, DA) when communication is prioritized ($\rho\rightarrow1$), but closely matches it when sensing dominates ($\rho\rightarrow0$) and frees up BSs for sensing.
\stepcounter{footnote}
\edef\qosfootnotenumber{\arabic{footnote}}
\begin{figure}[t!]
\centering
     % \begin{subfigure}[b]{0.67\columnwidth}
        \includegraphics[width=0.59
        \linewidth]{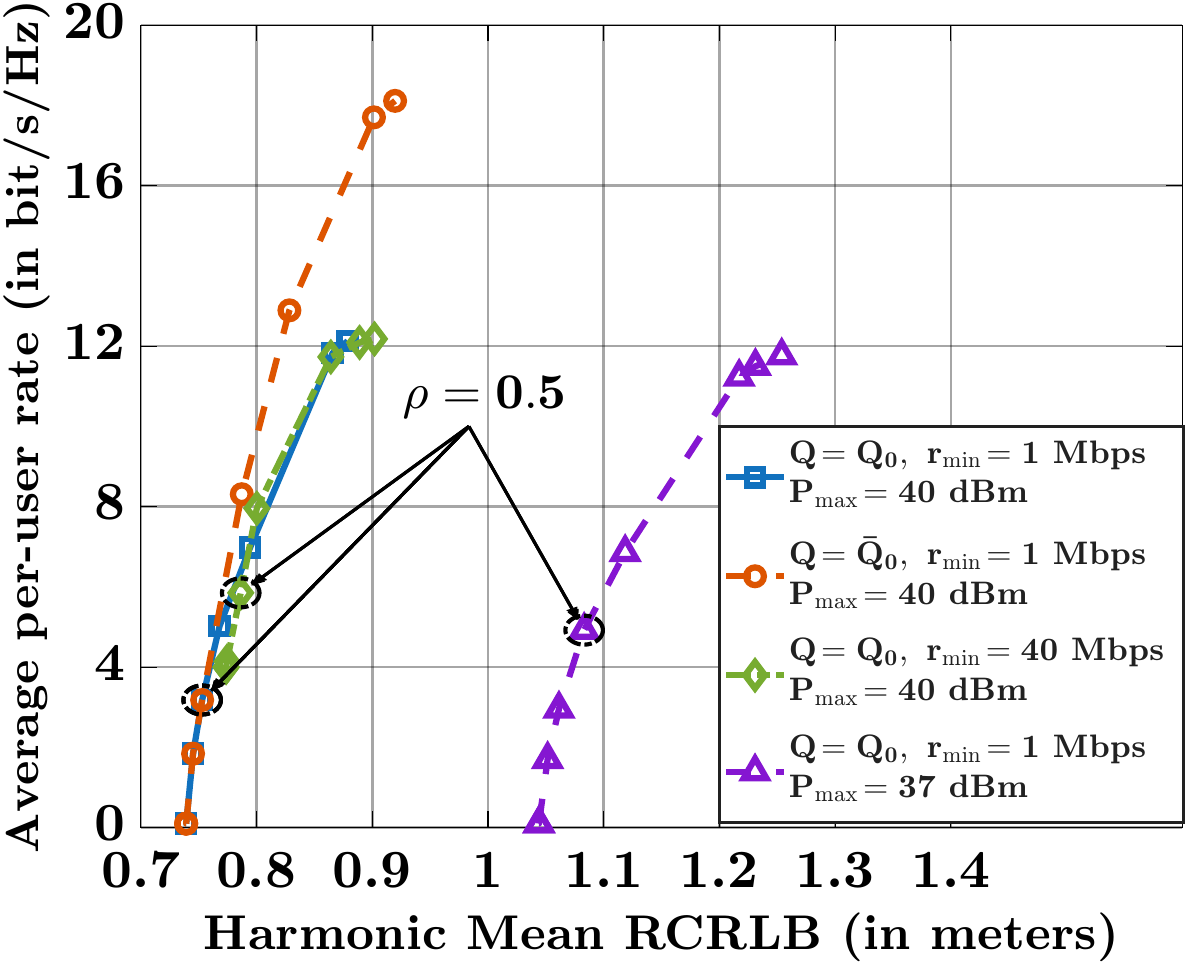}
    \vspace{-5pt}
    \caption[Analyzing performance bounds for the proposed DA under PS: varying $Q(t)$, $r_{\min}$, $P_{{\max}}$.]{Analyzing performance bounds for the proposed DA under PS: varying $Q(t)$, $r_{\min}$, $P_{{\max}}$. \protect\hyperlink{fn:qos}{\protect\footnotemark[\qosfootnotenumber]}}
     \vspace{-7pt}
    \label{fig:rQP_sweep}
  \end{figure}  
\begin{figure}[t!]
\centering
     % \begin{subfigure}[b]{0.67\columnwidth}
        \includegraphics[width=0.59
        \linewidth]{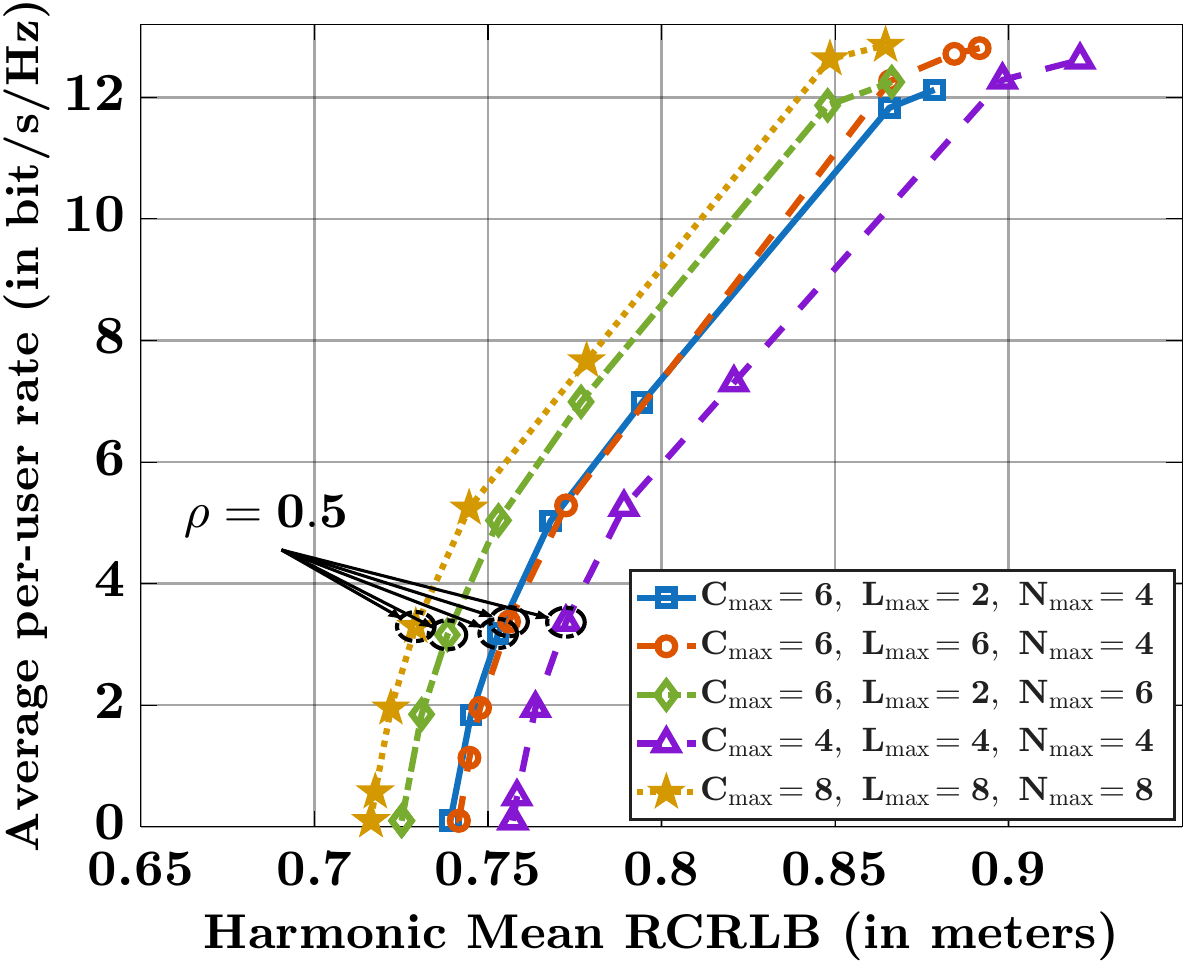}
        \vspace{-5pt}
    \caption{Analyzing the degrees of cooperation in the proposed DA scheme under PS: varying $C_{\max},L_{\max},N_{\max}$.\label{fig:Nmax_sweep}}
    \vspace{-11pt}
  \end{figure} 
% \subsection{Analyzing performance Bounds}
We consider the proposed scheme with a uniform backlog (with mean) $\bar{Q}_0(t) = 30$ kbits for all CUs in Fig.~\ref{fig:rQP_sweep} (default in~\cite{coop2,close2}). Avoiding early queue depletion for $2$ CUs at $\rho > 0.5$, yields up to a $5.7$~bps/Hz rate gain for only a $0.03$ m RCRLB penalty, further strengthening the work. Increasing $r_{\min}$ $40$-fold, rate raises from $0.1$ to $4$ bps/Hz at $\rho=0.1$ for just a $0.03$ m penalty. Halving the power budget to $P_{\max}=37$~dBm primarily penalizes two-way sensing by $0.35$~m. Interestingly, for $\rho\in[0.3,0.7]$ (below fair $\rho=0.7)$, rate is slightly better than the default case, indicating its much lower power demand under our sensing-prioritized utility scaling.
% Increasing $r_{\min}$ forty-fold ($40$~Mbps) raises the minimum rate at $\rho=0.1$ from $0.1$ to $4$~bps/Hz at the cost of only a $0.03$~m RCRLB increase, while converging to the default curve for $\rho \ge 0.7$ once the $4$~bps/Hz floor becomes inactive. Halving the power budget to $P_{\max}=37$~dBm primarily penalizes two-way sensing, increasing the RCRLB by $0.3\text{--}0.37$~m. Interestingly, for $\rho\in[0.3,0.7]$, $P_{\max}=37$~dBm achieves a slightly higher rate than $40$~dBm (e.g., $5$ vs.\ $3.2$~bps/Hz at $\rho=0.5$), as one-way communication links require significantly less power than two-way radar illumination.
% Fig.~\ref{fig:rQP_sweep} lists the same tradeoff curve for different queues $Q(t)$ ($t$ is omitted in the figure for brevity) and service guarantees ($r_{\min}$ and $P_{\max}$).  We also observe that when $r_{\min}$ is increased, we see a corresponding drop in sensing performance as communication improves. We can increase the transmit power to compensate for this drop as needed.%We see that if the queue had incoming data of the order of hundreds of kbps and not Mbps, correspondingly communication and sensing performance drops. --> I will be changing this curve to a more meaningful one
% % \subsection{Association Parameters and System Performance}
\footnotetext[\qosfootnotenumber]{%
\hypertarget{fn:qos}{$t$ is dropped from the figure for brevity.}%
}
% Finally, $N_{\max}, L_{\max}, C_{\max}$ determine the degree of `cooperation' in the system. In Fig.~\ref{fig:Nmax_sweep}, we analyze this on the sensing side. We observe a saturation of RCRLB at $N_{\max}=4$ (default case), similar to~\cite{coop2}. Even under a large sensing error ($\sigma_e=3$~m), the proposed scheme achieves an RCRLB of $\approx 2.8$~m, further confirming its robustness.increase as lesser BSs are available for each target under fixed $C_{\max}=6$. Conversely

In Fig.~\ref{fig:Nmax_sweep}, while adding $4$ BSs to each CU improves rate by $5.6$~bps/Hz at $\rho=0.7$ with a $0.09$~m RCRLB drop, adding $2$ BSs at targets enhances sensing by $0.02$~m while preserving rates, due to a fixed $C_{\max}\!=\!6$ and our sensing emphasis. Restricting BSs to $C_{\max}\!=\!L_{\max}\!=\!N_{\max}\!=\!4$ mimics the $P_{\max}=37$ dBm trend in Fig.~\ref{fig:rQP_sweep}: more impact on sensing by about $0.03$~m uniformly, whereas rate increases by $0.2$~bps/Hz for $\rho\in[0.5,0.7]$ due to higher $L_{\max}\!=\!4$ and earlier saturation in sensing gain. Finally, the $(8,8,8)$ case achieves the best $0.775$~m RCRLB and $7.9$ bps/Hz rate at a balanced $\rho=0.7$.

\section{Conclusion}
In this paper, we investigated the rate-RCRLB trade-off by optimizing communication and sensing beamformers for a cooperative multi-BS ISAC system under perfect and imperfect target knowledge. We proposed a dynamic strategy to solve a specific per-CU, per-target, and per-BS association problem. We proposed an iterative, SCA-based, low-complexity beamforming approach with closed-form updates to avoid costly SDP solvers. Simulation results analyzed the communication-sensing tradeoff for a variety of system configurations, thus establishing the superiority of the proposed scheme.
\appendices%Departing from existing works that rely on static topologies or single-BS associations,
% \section{Proof of Convexity of FIM $\boldsymbol{J}_m(t)$}
% \label{app:FIMconvexity}
% \begin{IEEEproof}
% $|f_{b,r,m}^{\mathrm{H}}(t)\boldsymbol{v}_{b,m}(t)|^2$ is convex in $\boldsymbol{v}_{b,m}(t)$. To prove 
% $\boldsymbol{J}_m(t)$ as a matrix valued function of $\boldsymbol{V}_m(t)$ is convex, we assume a fixed $\boldsymbol{y}\in\mathbb{R}^{2\times1}$ such that $\boldsymbol{y}^{\mathrm{T}}\boldsymbol{J}_m(t)\boldsymbol{y} = \left(1/\kappa_{\mathrm{nc}}\right)
% \sum_{b,r\in\mathcal{S}_m}
% |f_{b,r,m}^{\mathrm{H}}(t)\boldsymbol{v}_{b,m}(t)|^2 \boldsymbol{y}^{\mathrm{T}}\boldsymbol{z}_{b,r,m}\boldsymbol{z}^{\mathrm{T}}_{b,r,m}\boldsymbol{y}=\left(1/\kappa_{\mathrm{nc}}\right)
% \sum_{b,r\in\mathcal{S}_m}
% |f_{b,r,m}^{\mathrm{H}}(t)\boldsymbol{v}_{b,m}(t)|^2 |\boldsymbol{z}^{\mathrm{T}}_{b,r,m}\boldsymbol{y}|^2$. Non-negative weighted sum of convex functions yields convex $\boldsymbol{J}_m(t)$.
% \end{IEEEproof}

\bibliographystyle{IEEEtran}
% \bibliography{IEEEabrv}

\end{document}